\documentclass[sigplan,screen]{acmart}

\pdfoutput=1

\renewcommand\footnotetextcopyrightpermission[1]{} 

\acmISBN{} 
\acmDOI{} 
\startPage{1}

\setcopyright{none}

\usepackage{booktabs}
\usepackage{subcaption} 
\usepackage{multirow}
\usepackage{fancyvrb}

\begin{document}

\title[]{MP-CodeCheck: Evolving Logical Expression Code Anomaly Learning with Iterative Self-Supervision}

\author{Urs C. Muff}
\affiliation{ 
  \institution{Merly.ai} 
  \country{}
}
\email{urs.muff@merly.ai}

\author{Celine Lee}
\affiliation{
  \institution{Merly.ai, Cornell University} 
  \country{}
}
\email{celine.lee@merly.ai}  

\author{Paul Gottschlich}
\affiliation{
  \institution{Merly.ai} 
  \country{}
}
\email{paul.gottschlich@merly.ai}  

\author{Justin Gottschlich}
\affiliation{
  \institution{Merly.ai, University of Pennsylvania}
  \country{}
}
\email{justin.gottschlich@merly.ai}    

\begin{abstract} 

Machine programming (MP) is concerned with automating software development. According to studies, software engineers spend upwards of 50\% of their development time debugging software. To help accelerate debugging, we present \system{} (\systemshort{}). \systemshort{} is an MP system that attempts to identify anomalous code patterns within logical program expressions. In designing \systemshort{}, we developed two novel programming language representations, the formations of which are critical in its ability to exhaustively and efficiently process the billions of lines of code that are used in its self-supervised training. 

To quantify \systemshort{}'s performance, we compare it against ControlFlag, a state-of-the-art self-supervised code anomaly detection system; we find that \systemshort{} is more spatially and temporally efficient. We demonstrate \systemshort{}'s anomalous code detection capabilities by exercising it on a variety of open-source GitHub repositories and one proprietary code base. We also provide a brief qualitative study on some of the different classes of code anomalies that \systemshort{} can detect to provide an abbreviated insight into its capabilities.\footnote{Note: To ensure author anonymity, \systemshort{}'s GitHub repository has been withheld (and made private). The authors will open-source \systemshort{} upon completion of the review process.}

\end{abstract}



\newcommand{\system}{MP-CodeCheck}
\newcommand{\systemshort}{MPCC}

\maketitle
\pagestyle{plain}

\section{Introduction}
\label{sec:intro}
\begin{figure*}[htpb]
\begin{center}
\includegraphics[width=1.0\textwidth]{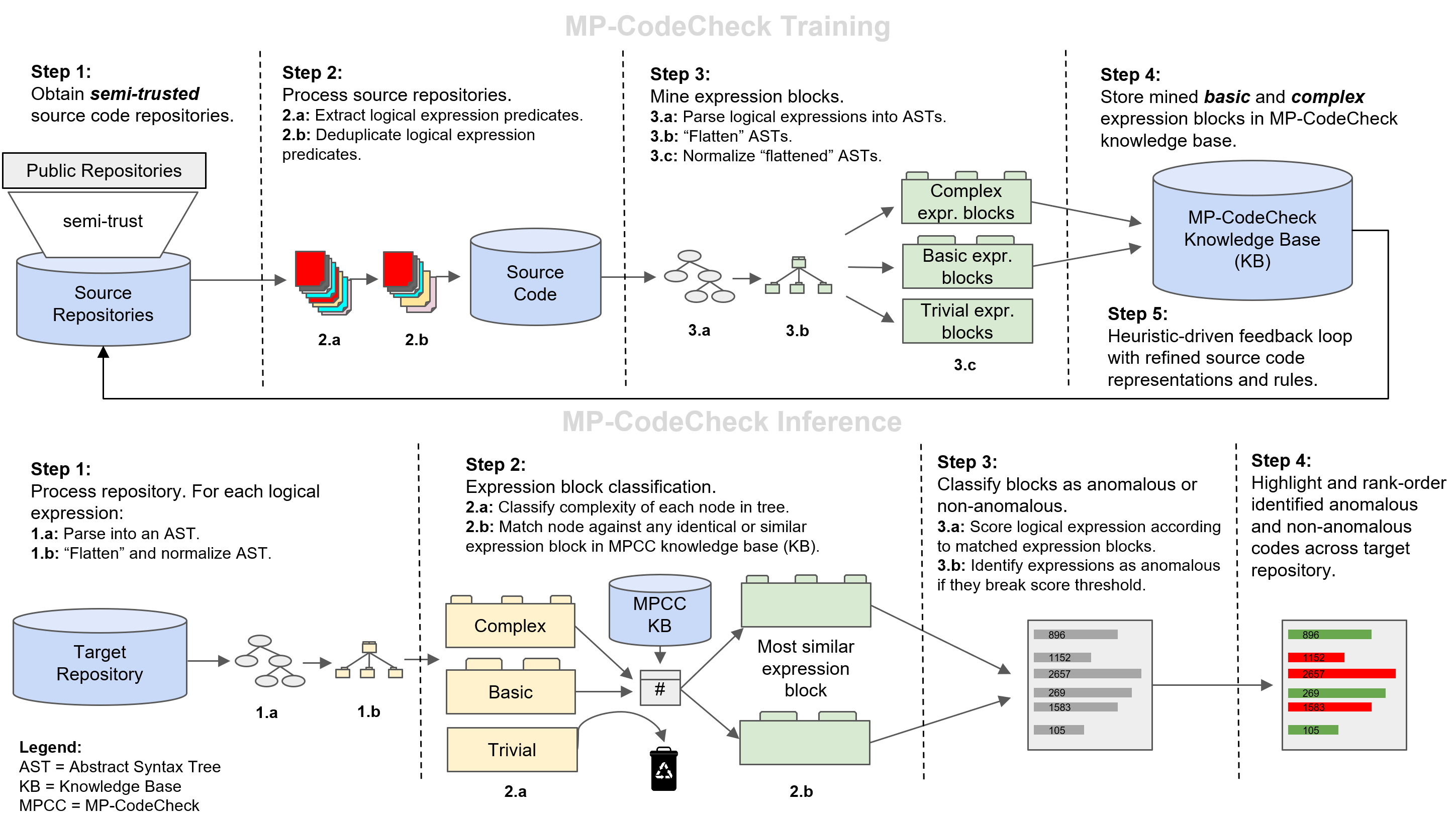}
\vspace{-0.15in}
\caption{System Overview of \system{} (\systemshort{}).}
\label{fig:system_overview}
\end{center}
\end{figure*}

Software debugging has been found to consume upwards of 50\% of all software development time~\cite{evansdebuggingtime}. \emph{Machine programming} (MP), the field concerned with the automation of all aspects of software development~\cite{gottschlich2018pillars}, has seen advances in software development related tasks such as code auto-completion~\cite{chen2021codex,svyatkovskiy:2019:pythia,guo2022learning}, code generation and program synthesis~\cite{balog:2017:iclr,becker:aiprogrammer:2021,kamil:verlifstencil:2016,mandal:2021:mlsys,yin-neubig-2017-syntactic,ling2016latent,trivedi2021learning,gulwani:2017:aplas,pmlr-v28-menon13,perelman:testsynth:2014,gulwani:2011:sigplan,solarlezama:sketching:2008,nye2019learning,yin:2018:emnlp,kulal:2019:neurips,ellis:dreamcoder:2021,murali2018neural,alphacode}, program transformation~\cite{gao:oopsla:2020,machines2020task-oriented} and repair~\cite{allamanis2021selfsupervised,li:2020:apr,Devlin2017SemanticCR,yasunaga:2020:sspr,pmlr-v139-berabi21a}, code similarity and recommendation~\cite{ye2020misim,luan:2019:aroma}, learned optimizations~\cite{patabandi:2021:tensoroptimization,marcus2020bao,Cen2020LearnedGC}, and performance regression testing~\cite{alam:zeroposregressions:2019,attariyan:xray:2012,li:pcatch:2018,song:statdebug:2014,nguyen:regression:2012}, amongst others. 

The latter few tasks support the MP goal of software adaptation, which focuses on evaluating or transforming higher-order program representations or legacy software programs to achieve certain characteristics (e.g., performance, security, etc.). An open challenge in software adaptation is reasoning about legacy software. In such code bases, software defects may arise from a number of issues. These include, but are not limited to, logical errors, poorly organized code, and technical debt. Even defects that do not have a clearly negative impact on a system may leave it susceptible to future vulnerabilities. Some of these defects manifest in weaknesses (or brittleness) in logical expressions. In an attempt to help remedy this, we present \textbf{\system{} (\systemshort{})}, a self-supervised, inventive, and adaptive MP system which detects anomalous logical expressions at the source code level. 


For the purposes of this paper, we trained \systemshort{} on over two billion lines of semi-trusted source code for common logical expression programming patterns, which we refer to as \textit{expression blocks}. Using the intuition that deviation from trusted programming techniques and paradigms can lead to potential erroneous programming, \systemshort{} uses mined expression blocks to identify anomalous code that make the program incorrect, prone to future bugs, or contain technical debt. For example, consider the following C/C++ example that (incorrectly) checks if a variable \texttt{x} is \texttt{NULL}:

\begin{footnotesize}
\begin{center}
\begin{Verbatim}[commandchars=\\\{\}]
  // malformed, but legal double NULL equality check
  \textcolor{blue}{if} (\textcolor{purple}{NULL} == x == \textcolor{purple}{NULL})
  \{
      throw std::runtime_error("x is NULL.");
  \}  
\end{Verbatim}
\end{center}
\end{footnotesize}

While this code snippet may look suspicious, it compiles without warning in the default configuration of Visual Studio 2022 and produces only a warning using GCC with compiler flag \verb|-Wpointer-arith|. However, the code's logical expression -- when considered holistically -- is erroneous. With \verb|x| set to \verb|NULL|, the execution of the \verb|if| statement will first check whether \verb|NULL == x|, which will evaluate to \verb|true|. Then the \verb|if| statement will continue its evaluation from left to right and compare \verb|true == NULL|, which will evaluate to \verb|false|. This will (incorrectly) cause the exception code that was meant to throw an exception when \verb|x = NULL| to be skipped. Given this, the system will likely dereference \texttt{x} at a later time causing an illegal memory access. \system{} was designed to identify these issues. In fact, \system{} found this anomaly, and others like it, in a large-scale production-quality software repository. 

In this paper, we make the following contributions:
\begin{enumerate}
    \item We present \system{} (\systemshort{}), a system that aims to identify anomalous logical expressions in code by incorporating several novel code representations and an iteratively-refined heuristic framework that guides \systemshort{}'s self-supervised engine.
    \item We present a spatial and temporal learning and inference performance comparison between \systemshort{} and ControlFlag~\cite{hasabnis2021maps} across 6,000 C/C++ repositories.
    \item We provide an analysis of \systemshort{}'s anomaly detection capabilities across ten GitHub repositories that are intentionally varied in size and lifetime.
\end{enumerate}

\section{Related Work}
There have been many recent works in the field of machine programming. In this section, we discuss some of the works that we have found are most relevant to \systemshort{}.

\subsection{Self-supervised Systems}
The emergence of self-supervised MP systems may be promising for large-scale machine learning, due to their ability to function on the enormous corpora of unlabeled open-sourced code training data. In the domain of natural language processing, large pre-trained language models~\cite{brown:gpt:2020, megatron:nvidiamicrosoft:2022, devlin-etal-2019-bert} have already shown to be powerful tools for few-shot learning in language processing tasks~\cite{schick-schutze-2021-exploiting,radford2019language} where task-specific or domain-specific labeled datasets may be unavailable. For code processing tasks, GitHub alone hosts over 46 million public software repositories, presenting one source of abundant but unlabeled code data. Recent self-supervised code processing systems such as OpenAI's Codex~\cite{chen2021codex}, Intel's ControlFlag~\cite{hasabnis2021maps}, Microsoft Research's BugLab~\cite{allamanis2021selfsupervised}, and the basis for DeepMind's AlphaCode~\cite{alphacode}, among others, have taken advantage of the vast amount of available code data to achieve impressive results for their respective tasks of code generation, idiosyncratic code pattern detection, bug detection and repair, and competitive programming. 

Codex~\cite{chen2021codex} and the underlying model for AlphaCode~\cite{alphacode} are both large transformer language models pre-trained on large amounts of unlabeled GitHub code. This model setup allows for Codex variants and AlphaCode to then be fine-tuned on some small dataset of domain-specific examples to perform a particular task; AlphaCode, for example, is fine-tuned on a competitive programming dataset to generate solutions to complex programming tasks.
ControlFlag~\cite{hasabnis2021maps} mined over one billion lines of unlabeled open-source C/C++ code for common and uncommon code patterns to detect idiosyncratic programming patterns. Once trained, it performs inference on user-supplied code and suggest corrections on anomalies it has found. 
BugLab~\cite{allamanis2021selfsupervised} takes an adversarial approach to learning software bug detection and repair by co-training a bug detection and repair model alongside a bug injection model such that the bug injection model learns to generate data from which the bug detection and repair model is trained. This boot-strapped technique does not require any external labeled training data. 


Other systems such as Snorkel~\cite{ratner:2019:vldb} combine weak supervision techniques to enable users to train state-of-the-art models without hand-labeling (much) training data.

\subsection{Program Repair / Program Synthesis}
Anomaly detection is closely related to other types of automated program reasoning such as program repair and program synthesis. These systems can be broken down into human-in-the-loop systems and closed loop systems.

\paragraph{Human-in-the-loop Systems} A primary challenge in automated program reasoning is code semantic understanding; an incorrect implementation with even minor syntactic differences (e.g., in C/C++ \verb|==| (equality), \verb|=| (assignment)) can produce drastically different software results. One way that previous systems have attempted to address this issue is to incorporate humans into the overall system. That is, humans can guide the MP system's choice and potentially reinforce its learning algorithm. 

Code recommendation systems are one such family of program reasoning systems. A \emph{code recommendation} system is an automated system that ingests data (in some form) and then recommends some code fragment that is meant to satisfy the supplied input. Microsoft IntelliSense~\cite{intellisense}, Tabnine~\cite{tabnine} and GitHub Copilot~\cite{copilot} are examples of commercially-deployed code completion suggestion tools. They can be integrated into interactive development environments (IDEs) and provide real-time suggestions for the programmer in the IDE. IntelliSense~\cite{intellisense} uses knowledge of programming language semantics to suggest possible variables, methods, fields, type parameters, constants and classes, among other completions, as the programmer edits code in the IDE. Tabnine~\cite{tabnine} and Copilot~\cite{copilot} train deep neural networks to learn from large corpora of source code to autocomplete whole lines or whole functions of code. Tabnine offers the option to fine-tune learned models on the user's (or user's team's) code to provide guidance aligned with the user's own code practices. Copilot adapts to edits made to its suggestions to match the user's coding style. Chaurasia and Mooney incorporate humans into a natural language-to-code generation system~\cite{chaurasia-mooney-2017-dialog} by using dialog to clarify user intent until it has enough information to produce correct code.

\textit{Code similarity systems} analyze code fragments and determine if they are semantically \emph{(i)} similar, \emph{(ii)} dissimilar, or \emph{(iii)} equivalent. Such systems can be used for a variety of purposes. For example, they can help to identify existing code intent and suggest alternatives that may be less brittle and easier to understand. The MISIM neural code similarity system uses a context-aware semantics structure to lift semantics from code syntax. It then scores the semantic similarity of any two semantics structures via an extensible learned scoring algorithm (usually in the form of a deep neural network)~\cite{ye2020misim}. Such a system could be used for many things, one being language-to-language transpilation. The Aroma code recommendation system performs code search using a novel \emph{simplified parse tree} (SPT), which elides away many syntactical details of the original code~\cite{luan:2019:aroma}. Using the SPT, the Aroma system aims to take incomplete code snippets and return (more) complete code snippets.

\begin{figure*}[htpb]
\begin{center}
\includegraphics[width=1.0\textwidth]{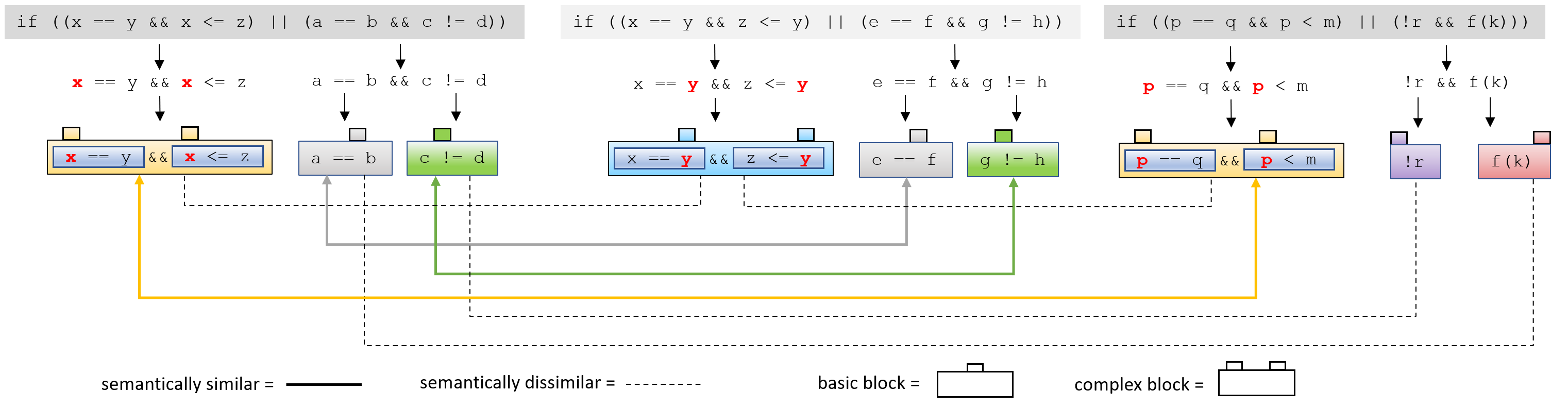}
\vspace{-0.15in}
\caption{Illustration of basic and complex expression blocks used in \system{}.}
\label{fig:example_blocks}
\end{center}
\end{figure*}

\paragraph{Closed Loop Systems} Some systems operate in an end-to-end manner to generate and modify code, without any human user feedback. Many program synthesis systems using sketching~\cite{solarlezama:sketching:2008}, inductive program synthesis ~\cite{gulwani:2017:aplas,pmlr-v28-menon13,perelman:testsynth:2014,feser:2015:sigplan, gulwani:2011:sigplan,lieberman:2001:book,mandal:2021:mlsys,balog:2017:iclr,pu:2018:pmlr,ellis:dreamcoder:2021}, and natural language descriptions~\cite{nye2019learning,wong:2006:naacl,kate:2005:ncai,quirk-etal-2015-language} operate in this closed loop manner to generate code satisfying some input specification.

One explored application of program transformation is \emph{transpilation}, which is a technique in translating code from one programming language to another. \citet{kamil:verlifstencil:2016} demonstrate a technique to lift low-level Fortran code to a high-level predicate language summary and lowering it back down into Halide code to achieve performance speedups. \citet{chen2018treetotree} also develop a tree-to-tree model to translate programs from one programming language to another.

\textit{Automated program repair (APR)}~\cite{le_goues_2019_apr} is the task of automatically repairing software to reduce the work of human engineers while maintaining program usability and avoiding software regression. BugLab~\cite{allamanis2021selfsupervised} is one example APR system that co-trains a bug detection and repair model alongside a bug injection model such that as the bug injection model learns to generate harder-to-find bugs, the bug detection and repair model learns to find and repair harder to find bugs. \citet{li:2020:apr} present a technique that uses prior bug fixes and surrounding code context to modify a buggy program's abstract syntax tree. Some APR works have also leveraged natural language to model reasoning about automated repair, such as \citet{yasunaga:2020:sspr} who use a buggy program's diagnostic feedback error message to localize then generate a repaired version of the erroneous line in the software source code. TFix~\cite{pmlr-v139-berabi21a} is another end-to-end text-to-text system that fixes buggy code without labels by pre-training a model on natural language then fine-tuning it on generating code fixes. Hoppity~\cite{Dinella2020HOPPITY} approaches automatic program debugging by learning a sequence of graph transformations over a buggy program represented as a graph structure.

These systems, however, focus on syntactic bugs that cause incorrect or failed program execution. \textit{Semantic program repair} is the task of fixing non-syntactic bugs that cause program behavior divergent from what the programmer intended. \citet{Devlin2017SemanticCR} propose an approach for automatic semantic program repair without access to the code's intended correct behavior at either training or test time: the system first proposes many potential bug repairs then scores them using a learned neuro-symbolic network, outputting the highest-scoring candidate as the solution.

\section{MP-CodeCheck System Design}
\label{sec:systemdesign}

\systemshort{}'s system overview is shown in Figure~\ref{fig:system_overview}. At the highest level, \systemshort{} can be thought of as a code anomaly detection system, similar to ControlFlag \cite{hasabnis2021maps}. However, \systemshort{} has at least two fundamental design departures from such prior works. First, it uses novel code representations that help its anomalous code detection engine. Second, it uses an iterative, programmatic heuristic to guide its self-supervised engine. As described in Section~\ref{sec:quantitative}, we have found that these design elements can reduce computational overhead (see Section~\ref{sec:quantitative}) as well as reducing false positives (see Section~\ref{sec:qualitative}). Moreover, by using these design elements together, \systemshort{} can manually or automatically be augmented to fit different programming languages, development environments, or stylistic constraints. Without these capabilities, it can be challenging or impossible to achieve similar customization (and debugging augmentation) when using a machine learning-only approach. This is especially true for systems that do not provide insight into their underlying mechanics (e.g., ControlFlag's string pattern matching algorithm). We describe both of these design elements in this section.

\subsection{Novel Code Representations}

\systemshort{} combines existing and novel code representations for its predicate expression classification system. Most of \systemshort{}'s internal code manipulation uses representations that are implemented using various graph structures, generally in the form of a tree (e.g., abstract syntax tree (ASTs), flattened non-binary tree, etc.). We have designed two new representations to enhance \systemshort{}'s ability to reason about the semantic properties of logical expressions: \emph{basic expression blocks} and \emph{complex expression blocks}. We describe them as follows.

\subsubsection{Basic and Complex Expression Blocks}

A novelty in \systemshort{} code representation is in its utilization of \textit{basic} and \textit{complex expression blocks}. The purpose of these blocks is to help the system reason about the semantics of logical expressions that may be asymmetrically spread across multiple logical operations in the same control structure. 
The formation of semantically rich compound expressions -- often found in complex expression blocks -- has helped \systemshort{} distill semantically complex expressions that are both nominal (i.e., non-anomalies) and anomalous. This distillation helps \systemshort{} identify more true positives as well as assisting in reducing false positives. We formally define basic and complex expression blocks as follows.

A \emph{basic expression block} is a predicate expression that contains no logical conjunction operators (i.e., it contains no logical-ANDs, logical-ORs). Every predicate expression can be divided into its basic expression blocks as shown in Figure~\ref{fig:example_blocks}. A \emph{complex expression block} is a composition of at least two basic expression blocks, normalized across variable (identifier) names. The intuition behind a complex expression block (or \emph{complex block}) is that program semantics are often encapsulated across multiple, disjoint predicates in a control structure. For example, in Figure~\ref{fig:example_blocks}, for each of the complex blocks, multiple predicates with a single, shared identifier must be satisfied for the complete logical expression to be true (e.g., \texttt{x} in the left-most yellow complex block, \texttt{y} in the blue complex block, and \texttt{p} in the right-most yellow complex block). These shared identifiers often embed semantic relationships between the basic expression blocks. To illustrate this concretely, consider the following common programming idiom to ensure a variable is within a minimum and maximum bounds:

\begin{footnotesize}
\begin{center}
\begin{Verbatim}[commandchars=\\\{\}]
        \textcolor{blue}{if} (min \textcolor{blue}{<} x \textcolor{blue}{&&} x \textcolor{blue}{<} max) \{ ... \}
\end{Verbatim}
\end{center}
\end{footnotesize}

The identifier \texttt{x} is present in both of the basic expression blocks (i.e., \texttt{min < x} and \texttt{x < max}). As such, if used with \systemshort{} a new complex expression block that conjoins both \texttt{min < x} and \texttt{x < max} would be constructed. When applied during inference, if \systemshort{} constructed a complex expression block that had such a signature (after being normalized), it would flag the block as non-anomalous as it would have learned this expression is nominal.

Our experience with \systemshort{} is that this use of complex expression blocks to capture compound programmatic semantics helps identify more complex programming anomalies, while simultaneously flagging nominal complex code structures that are common, such as the minimum and maximum expression discussed here.

\subsection{Programmatic Evolution of Self-Supervision}

A second novelty of \systemshort{} is in its machine-driven guidance on the evolution of its heuristic-based rules. This evolutionary process is a key aspect in the development of the core elements of the system as well as in improving the quality of the results. This step is captured in Step $5$ in \systemshort{}'s Training in Figure~\ref{fig:system_overview}. In this process, the knowledge that the self-supervised system learns, the common and uncommon control structure patterns, is used to inform human programmers in their construction of heuristics, representations, and rules that further improve the self-supervised learning. This iteration was done over a dozen times during the construction of \systemshort{}. 

A key reason for why this approach is critical to \systemshort{} is that in our experience, isolated self-supervision is generally insufficient to learn complex and nuanced concepts in programming languages; on its own, self-supervision often produces a large number of false positives. Instead, a heuristic-driven and iterative learning approach, that utilizes subject matter experts for reinforcement learning, helps to properly guide \systemshort{} through nuanced and unknown software design patterns, whether infrequent but correct, or frequent and incorrect.

Moreover, although it is not captured in this paper, in our early experiments, which did not use this human-in-the-loop iterative feedback loop, \systemshort{}'s false positive rate for code anomalies exceeded $\approx{90}\%$. When combining humans and machines in \systemshort{}'s design, it has reached $\approx{31\%}$ false positive rate (see Table~\ref{table:repo_metrics}), a $3\times$ reduction in false positives.

\begin{figure}[]
\begin{center}
\includegraphics[width=0.45\textwidth]{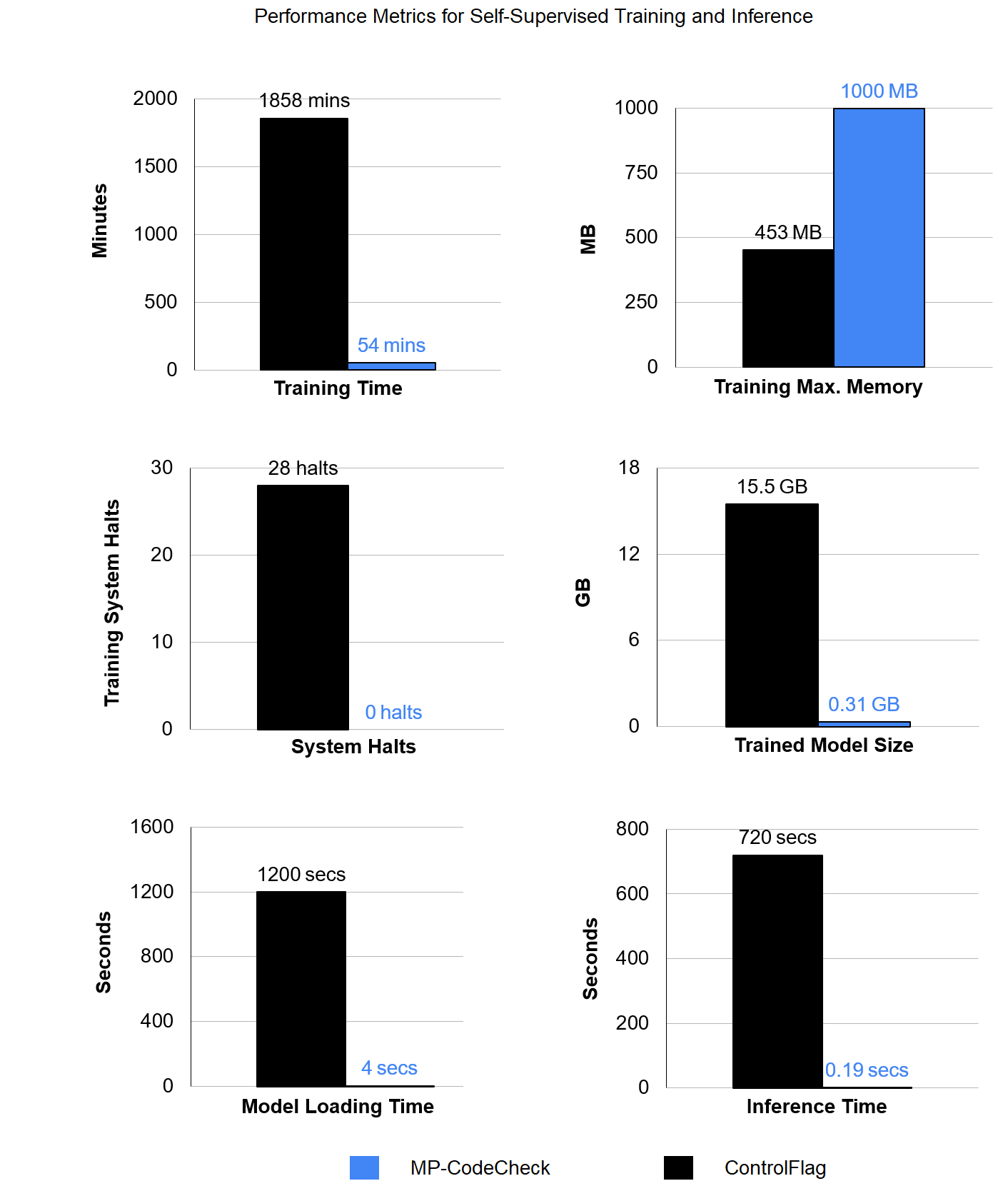}
\caption{Performance comparison: \system{} vs. ControlFlag in terms of training time, model size, model loading time, inference time, and reliability (i.e., number of halts).\protect\footnotemark}
\label{fig:performance}
\end{center}
\end{figure}
\footnotetext{All experiments run on the same system: OS: 64-bit Windows 11 Home; Processor: 11th Gen Intel(R) Core (TM) i7@ 2.80GHz; Ram: 16.0 GB; Drive: 1TB SSD}

\begin{table*}[]
\centering
\begin{tabular}{||r l l l l l l||} 
 \hline
  Repository & Size (MB) & Est. Year & \# of Anoms & \# of Expr. & Anoms Per Expr. & Top 20 False Positive \% \\ [0.5ex] 
 \hline\hline
 git/git                & 40.07     & 2008 & 41     & 31,298    & 0.131\%   & 55\% \\
 curl/CURL              & 15.92     & 2010 & 16     & 13,983    & 0.114\%   & \textbf{\color{red}{56.25\%}} \\  
 iovisor/bcc            & 15.14     & 2015 & 7      & 3,607     & 0.194\%   & 42.86\% \\ 
 netdata/netdata        & 44.26     & 2013 & 48     & 26,092    & 0.184\%   & 30\% \\ 
 php/php-src            & 125.27    & 2011 & 76     & 47,994    & 0.158\%   & 10\% \\ 
 proprietary*           & \textbf{\color{black}{196.06}}    & \textbf{\color{black}{1999}} & 66     & 12,948    & 0.510\%   & 20\% \\ 
 qemu/qemu              & 116.78    & 2012 & \textbf{\color{red}{122}}    & \textbf{\color{red}{79,050}} & 0.154\%   & 35\% \\ 
 raspberrypi/pico-sdk   & 7.25      & \textbf{\color{black}{2021}} & 22     & 1,226     & \textbf{\color{red}{1.79\%}}    & 27.28\% \\  
 shakevsky/keybuster    & \textbf{\color{black}{2.47}}      & 2020 & \textbf{\color{blue}{2}}      & \textbf{\color{blue}{393}}       & 0.509\%   & \textbf{\color{blue}{0\%}} \\ 
 ventoy/Ventoy          & 112.2     & 2020 & 22     & 17,316    & \textbf{\color{blue}{0.127\%}}   & 35.29\% \\ 
 \hline
 \textbf{\emph{Averages}}  & 67.4 & N/A & 42.2 & 23,390 & 0.387\% & \textbf{31.16\% FP Rate} \\
 \hline\hline
  Repository & Size (MB) & Est. Year & \# of Anoms & \# of Expr. & Anoms Per Expr. & Top 20 False Positive \% \\ 
 \hline\hline
\end{tabular}
\caption{Results of \systemshort{}'s Inference on 10 Repositories Ranging in Size and Year of Establishment.}
\label{table:repo_metrics}
\end{table*}

\section{Quantitative Results}
\label{sec:quantitative}

In this section, we present quantitative results on the \system{} system. These include \textit{(i)} performance metrics of \systemshort{} compared to ControlFlag and \textit{(ii)} \systemshort{}'s inference accuracy metrics with respect to false positive rates across several open-source GitHub repositories (see Table~\ref{table:repo_metrics}).~\footnote{We tried to compare \systemshort{}'s inference results to those from ControlFlag, but were unable to due to computational tractability limitations in ControlFlag's open-source system.}

\subsection{Computational Performance Metrics}

Figure~\ref{fig:performance} details the performance results of our experimental study comparing \systemshort{} to ControlFlag. We trained both systems on the open-source code data recommended by the ControlFlag repository's README\footnotemark\footnotetext{https://github.com/IntelLabs/control-flag/blob/master/README.md}, which consists of 6,000 repositories containing ~1.1 billion lines of C code with some minor C++-specific code ($< 5\%$). The training metrics we gathered include training time (in minutes), maximum memory utilized during training (in MB), number of software halts during training, and resulting trained model size (in GB). For inference, we ran both systems on the open-source GitHub Load Balancer \footnotemark\footnotetext{https://github.com/github/glb-director} repository and collected metrics on model loading and inference time across the entire repository.

\subsection{\systemshort{}'s Basic and Complex Expression Blocks for Computational and Spatial Efficiency}

\systemshort{}'s basic and complex expression block representations are core components of its efficient training, inference, and model size. For example, ControlFlag uses a syntax-driven trie (a type of k-ary search tree) for its training and inference. This syntax-driven trie stores and compares each individual syntax letter of an expression in trie form to numerous other expression tries from both anomalous and non-anomalous clusters. This process is performed iteratively, per syntax element (i.e., letter), until an identical match, or no match at all, is found. While mathematically sound, this approach suffers from search space growth with an upper bound that is exponentially proportional to the size of the alphabet used in the associated search grammar.

On the other hand, for training and inference expression matching, \systemshort{} uses a restricted search that only consists of basic and complex expression blocks that have already been learned as potential matches. In total, these blocks constitute less than $70,000$ unique entries for the one billion lines of code used for training. Moreover, when compared to a syntax trie counterpart, the number of unique basic and complex expression blocks has a relatively small numerical bound. This is due to two reasons. First, each expression block is reduced from its original syntax into a minimized normalized form, which reduces its spatial footprint and enables divergent syntax expressions to converge into semantically equivalent normalized (and minimized) block form. Second, the expression blocks are limited to those that are accepted design patterns found in the semi-trusted training repositories, thus limiting the total number of unique blocks. The computational divergence of these two search approaches leads to notable overall system performance differences between \systemshort{} and ControlFlag as shown in Figure~\ref{fig:performance}.\footnote{Note that \systemshort{}'s utilization of ${\approx{2\times}}$ more available memory during training than ControlFlag results in an additional computational efficiency over ControlFlag, as \systemshort{} is able to more fully exercise the available hardware.}

\subsection{System Accuracy}

In Table~\ref{table:repo_metrics}, we show our experimental results for \systemshort{} inference accuracy. We tested \systemshort{} on ten repositories with an intentional variation in repository size and year of establishment. For each repository, \systemshort{} parses all C and C++ source files for control expressions then classifies each as anomalous or non-anomalous according to the scoring system explained in Section~\ref{sec:systemdesign}. In our experiments, we set the anomaly threshold to $1000$: if a logical expression is assigned a complexity score greater than $1000$, then \systemshort{} flags it as an anomaly. We then manually inspected 20 anomalies with the highest anomaly complexity scores per repository to determine whether the flagged anomaly is not in fact an anomaly (i.e., it is a false positive). 

Across the ten repositories that we inspected, we found a false positive rate of $31.16\%$. Anomalies are marked as false positives if they do not introduce excess technical debt. That is, most anomalies that are \textit{not} false positives (i.e., true positives) possess one or more of the following features:

\begin{enumerate}
    \item They use improper pointer checking practices.
    \item They have unclear arithmetic operations. 
    \item They have unclear and inconsistent type casting.
    \item They have many connected but disjoint predicates (i.e. each predicate performs a check on a different variable) for error checking.
    \item They have inefficient usage of logical operators.
    \item They have over- or under-utilization of parentheses.
    \item They are C++ specific operations.
    \item They perform arithmetic and Boolean operations on the same variable.
    \item They are potential bugs.
\end{enumerate} 

\begin{figure*}[]
\begin{center}
\includegraphics[width=1.0\textwidth]{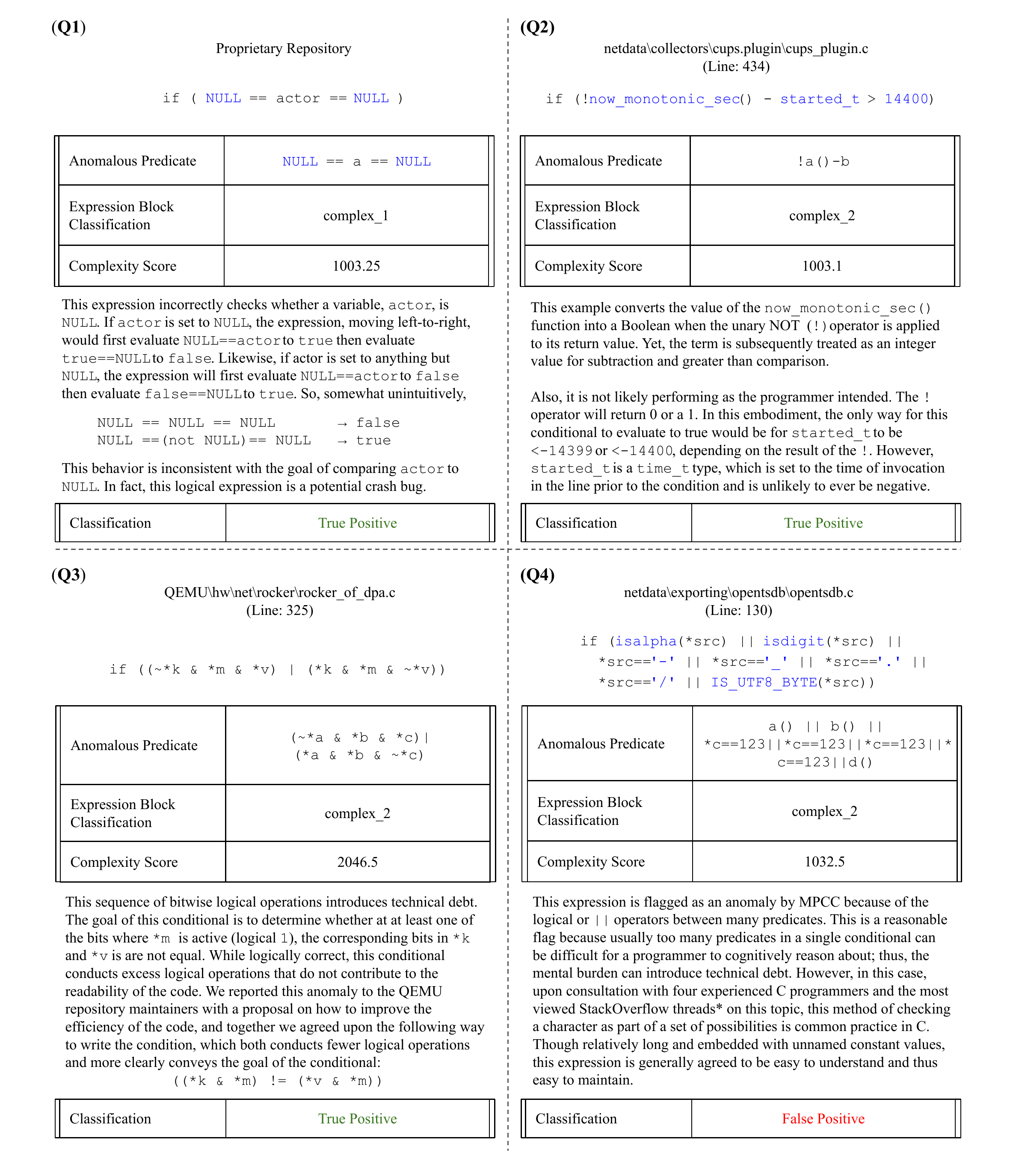}
\caption{Four Qualitatively Analyzed Examples of Anomalous Logical Code Expressions as Detected by MPCC.~\protect\footnotemark}
\label{fig:anecdotal_anomalies}
\end{center}
\end{figure*}
\footnotetext{\url{https://stackoverflow.com/questions/39150884/is-there-a-shorter-way-to-write-compound-if-conditions/39151002##39151002}, \url{https://stackoverflow.com/questions/16644906/how-to-check-if-a-string-is-a-number/16644949##16644949}, \url{https://stackoverflow.com/questions/13214506/shorthand-for-checking-for-equality-to-multiple-possibilities}}

\section{Qualitative Results}
\label{sec:qualitative}

Figure~\ref{fig:anecdotal_anomalies} presents qualitative results on the \system{} system by analyzing four flagged anomalies in detail. Each of the four flagged anomalies, shown in quadrants (Q1)-(Q4) of the figure, were flagged by \systemshort{} because the programming patterns that they exhibit, as represented by \systemshort{}'s novel code representation structure, were uncommon or unseen before in the C code on which \systemshort{} was trained. The first three anecdotal examples (Q1)-(Q3) are true positives, or true anomalies. The first example (Q1) is the same one as introduced in the Introduction (Section~\ref{sec:intro}). The last anecdotal example (Q4) is a false positive: \systemshort{} flagged it as an anomaly, but upon manual inspection, it is actually non-anomalous. We detail our analysis of each anecdotal example in their corresponding quadrants in Figure~\ref{fig:anecdotal_anomalies}.

\section{Conclusion}

In this paper, we introduced \system{} (\systemshort{}), a self-supervised anomaly detection system for logical expressions. Our early evidence seems to demonstrate that \systemshort{} can assist in one of the more painstaking aspects of software development, debugging, by identifying anomalies in even hardened production-quality code. We also demonstrated that \systemshort{} is more temporally and spatial efficient than ControlFlag, a state-of-the-art self-supervised anomaly detection that also identifies anomalies in logical code expressions. Moreover, our early results across ten high-quality code repositories, rates \systemshort{} with a false positive rate of $\approx{31\%}$. In conducting our experimentation with \systemshort{} on these repositories, we identified what we believe are not only anomalies with technical debt, but also more serious ones such as security vulnerabilities and illegal memory accesses (e.g., crash bugs).

\section{Broader Impact}

\system{} uses semi-trust to obtain its training data, which means that \systemshort{} could be susceptible to unintentionally learning from untrustworthy code. For example, if an adversarial attacker were to provide training code data with many instances of malicious programming patterns, the resulting \systemshort{} model would likely exhibit many false negatives and false positives. That is, it may flag non-anomalous code as anomalous and anomalous code as non-anomalous. 

Another broader impact to consider is \systemshort{}'s demand for computational resources. \systemshort{} obtains its knowledge by mining billions of lines of code, which demands nontrivial amounts of computation, albeit much less than any deep learning based counterpart. This demand for computation places some amount of strain on both the environment, which can contribute to climate-related issues, and the global supply chain, which can contribute to economic issues. 

As a machine programming system, \systemshort{} has the objective of reducing or eliminating some burdens of software development. This may seem to potentially reduce demand for software engineers. However, we believe that the tasks that \systemshort{} alleviates are tasks that, despite being critical to software robustness, are principally only understood deeply enough to be done by a small minority of existing software developers. These tasks also take up a sizable chunk of software development time, whether in hunting down and fixing code anomalies or in fixing bugs manifested by unremedied code anomalies. We believe that by automating anomaly detection, \systemshort{} would increase productivity of software developers and subsequently open up time for more creative tasks such as algorithm design and entrepreneurship, making robust software engineering more accessible to even more software developers.

\bibliography{ms}

\appendix

\end{document}